\documentclass[doublecol]{epl2}

\usepackage{graphicx}
\usepackage{eurosym}
\def\junk{{\it junk }}

\title{Correlations and Omori law in Spamming}

\author{Massimo Pica Ciamarra\inst{1,2} \and Antonio Coniglio\inst{2} \and Lucilla de Arcangelis\inst{1}}
\shortauthor{Massimo Pica Ciamarra, \and Antonio Coniglio, \and Lucilla de Arcangelis}

\institute{
 \inst{1} Department of Information Engineering and CNISM, Second University of Naples, 81031 Aversa (CE), Italy\\
 \inst{2} Department of Physical Sciences, University of Naples ”Federico II”, 80125 Napoli, Italy, INFN and Coherentia
}


\abstract{
The most costly and annoying characteristic of the e-mail communication system is the large number of 
unsolicited commercial e-mails, known as spams, that are continuously received. 
Via the investigation of the statistical properties of the spam delivering intertimes, we show that
spams delivered to a given recipient are time correlated: if the intertime between two consecutive spams 
is small (large), then the next spam will most probably arrive after a small (large) intertime. 
Spam temporal correlations are reproduced by a numerical model based on the random superposition of spam sequences, each one described by the Omori law. 
This and other experimental findings suggest that statistical approaches may be used to infer how spammers operate.
}

\begin{document}

\maketitle

\section{Introduction}
Quoting from Ref.~\cite{EU}, a press release of the European Union:
``The proliferation of unsolicited commercial e-mail, or `spam', has reached a point where it creates a major problem for the development of e-commerce and the Information Society. Businesses and individuals spend an increasing amount of time and money simply to clean up e-mailboxes. The loss in productivity for EU businesses has been estimated at $2.5$ billion \euro~ for 2002. [$\ldots$]  Spam has the potential of destroying some of the major benefits brought about by services such as e-mail and SMS.''

Spams, defined as undesired commercial e-mails, are estimated to be $70-80\%$ of all e-mails~\cite{antispam}, 
as everybody has probably noticed when opening his e-mail box. Such a large number is explained by considering that the daily earning of a spammer is proportional to the number of spams sent. To reduce the nuisance caused by spams, an enormous effort has been devoted to design efficient spam filters (see~\cite{goodman} and references therein), able to quickly discriminate between a spam and a legitimate e-mail. Much less effort has been devoted to the problem of understanding how spammers operate, which is the crucial information required to fight spammers at the source. 
In this paper,  we present a statistical analysis of the spamming process, which may help unveil how spammers operate.

\section{Dataset}
Our analysis has been made possible by modern antispam filters, able to discriminate with good accuracy between legitimate e-mails and spams. These filters can be configured in such a way that spams are not erased, but collected in an appropriate folder: we call this folder the \junk folder. We have considered four \junk folders, $J_1,J_2,J_3, J_4$, belonging to four academic e-mail accounts of our university (domain `na.infn.it'). The folders are created by the antispam filter ``Sophos'', and contain respectively $16\cdot10^4$, $27\cdot10^3$, $21\cdot10^3$,  and $7\cdot10^3$ spams. 
For comparison, we have also considered one standard inbox folder, $I$, containing $4\cdot10^3$ legitimate e-mails. The popularity of the four accounts we have considered among spammers varies, as the mean intertime between two consecutive spams is, respectively,  $300s$, $700s$, $1100s$ and $870s$ seconds.
For each e-mail, we have determined the time of arrival and the geographical location of the sender. The delivering time of an e-mail $t_i$ is registered by the incoming mail server. In order to obtain an estimate of the error on $t_i$, we set-up a script to send at a regular interval, $t_{\rm delay}$, e-mails from an account $Bob$ (based in the USA) to a different account, $Alice$, (based in Italy). The intertime between two consecutive e-mails delivered to $Alice$ is not constant and equal to $t_{\rm delay}$, but fluctuates. The typical size of these fluctuations (which may depend on the internet routing) is $10$s. This value is our estimate for the error on the delivering times. The geographical location of the sender is determined from the IP address of the sender~\cite{ip}, which is recorded in the envelop which complements any e-mail.

\section{Data analysis}
\begin{figure}[t!!]
\begin{center}
\includegraphics*[scale=0.31]{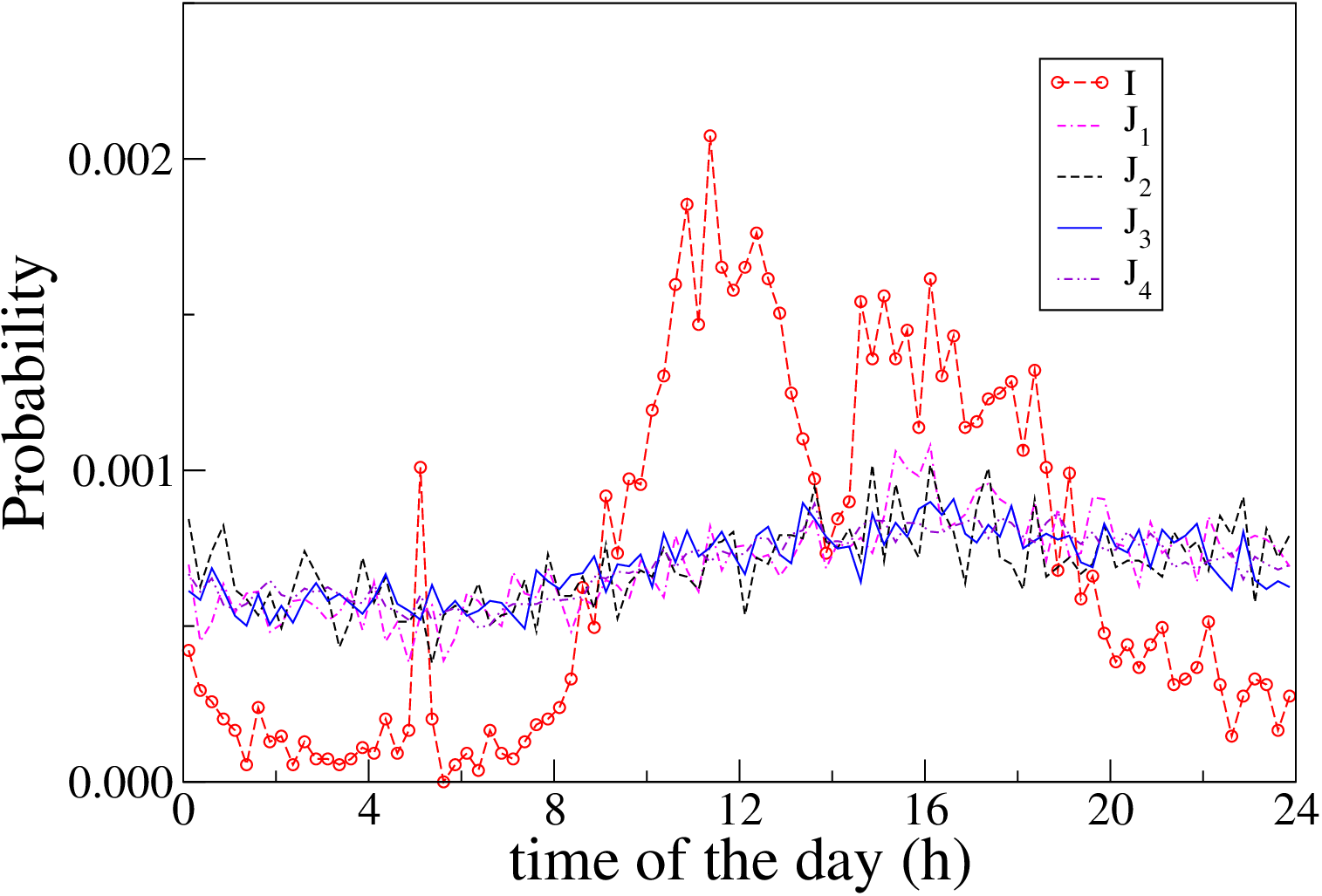}
\end{center}
\caption{\label{fig:mindist} (color online) Daily dependence of the probability of receiving an e-mail. $I$ indicates the probability of receiving a regular e-mail, which is highly structured. $J_i$ indicates the probability of receiving a spam, which on the contrary exhibits small oscillations during the day. The index $i = 1,2,3,4$ identifies the investigated datasets.
}
\end{figure}

The investigation of the e-mails received by the incoming mail server~\cite{Gomes}, as shown in Fig.~\ref{fig:mindist}, reveals that spams and legitimate e-mails have different statistical features. For instance, regular e-mails traffic has a temporal modulation which clearly reproduces humans activity (small activity during the night and at lunch time),
whereas spams appear to be insensitive to it. This is a clear signature of the different mode of operation of the e-mail senders, ``whereas legitimate e-mail transmissions are driven by social bilateral relationships, spam transmissions are a unilateral spammer-driven action''~\cite{Gomes}. 

\begin{figure}[t!!]
\begin{center}
\includegraphics*[scale=0.31]{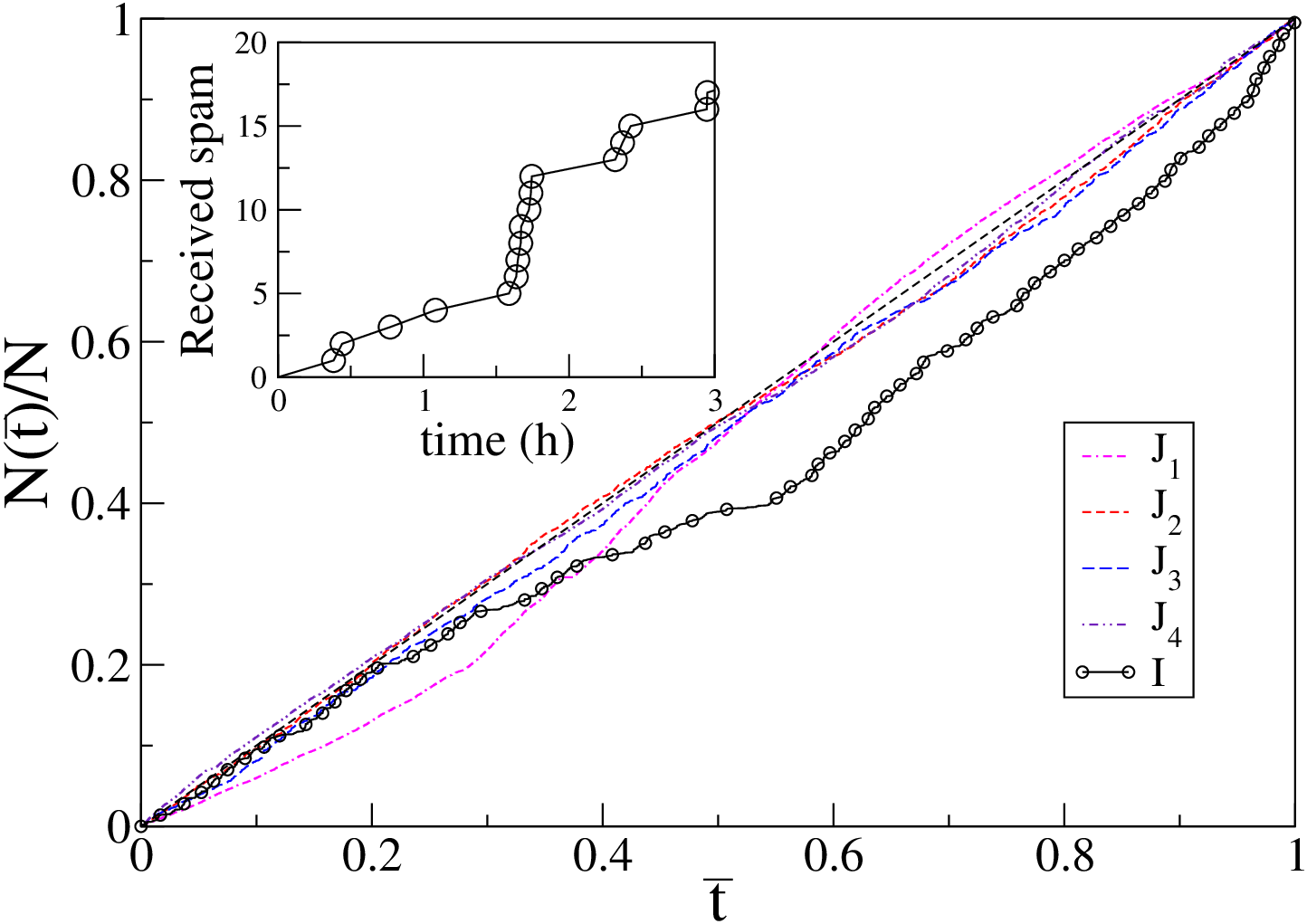}
\end{center}
\caption{\label{fig:rate} (color online) The cumulated number of delivered e-mails $N(t)$ increases linearly in time. In order to rescale the data of different datasets, here we plot $N(t)/N$, where $N$ is the number of e-mails of the dataset, as a function of the rescaled time $\overline{t}= (t-t_1)/(t_N-t_1)$, where $t_1$ and $t_N$ are the time of arrival of the first and of the last e-mail of the dataset. For a stochastic process at constant rate $N(t)$ increases linearly in time. The inset shows how the number of received spams increases in a temporal window of $3h$, and shows the existence of bursts.
}
\end{figure}

As shown in Fig.~\ref{fig:rate}, spams are received at constant rate; yet the time series of the arrival times $t_i$ is characterized by bursts, short temporal periods during which many correlated spams are received. Here we suggest that the origin of these bursts lies in the use of peer-to-peer (i.e. decentralized) network of infected computers to send spams, known as botnets.


\begin{figure}[t]
\begin{center}
\includegraphics*[scale=0.31]{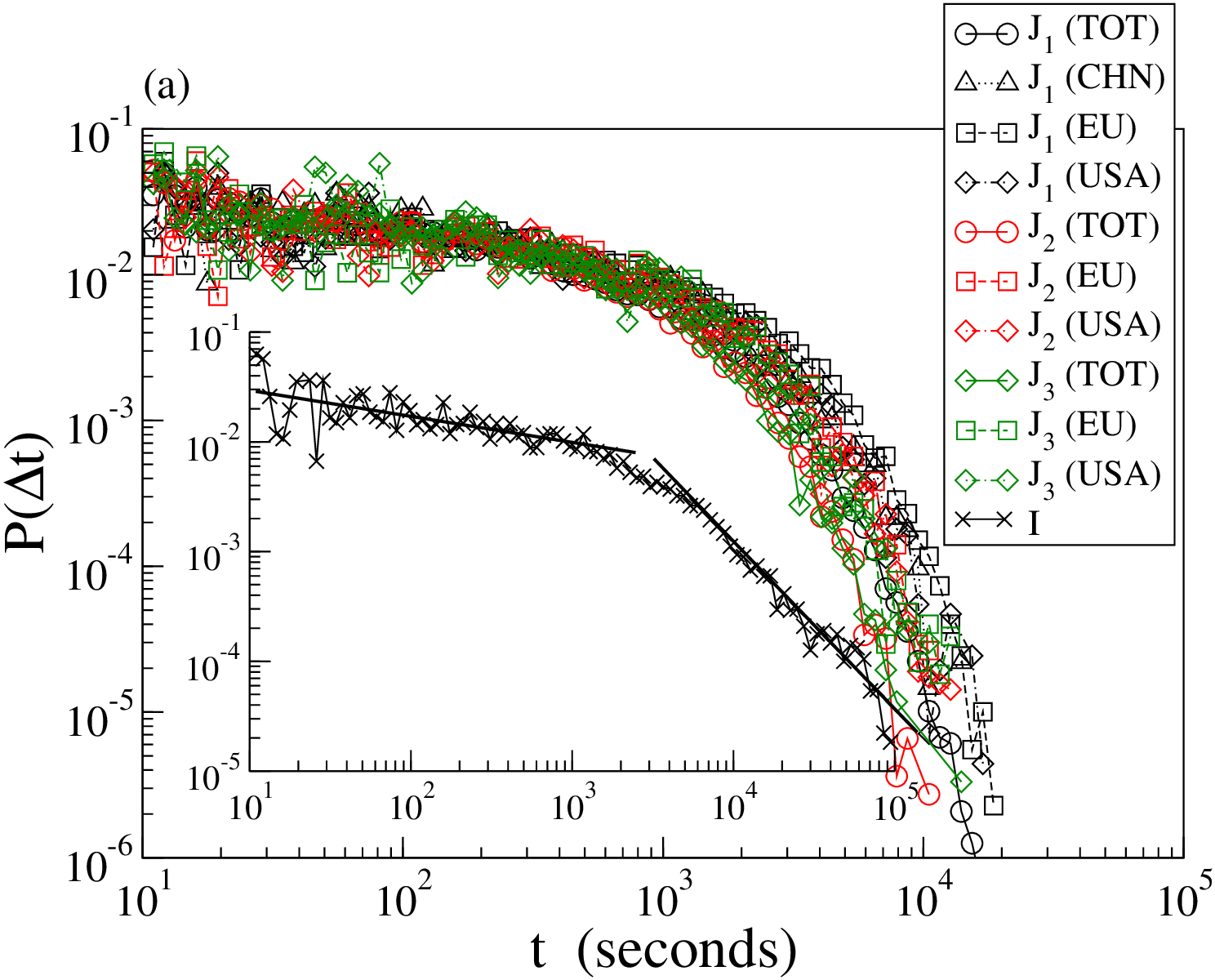}
\includegraphics*[scale=0.31]{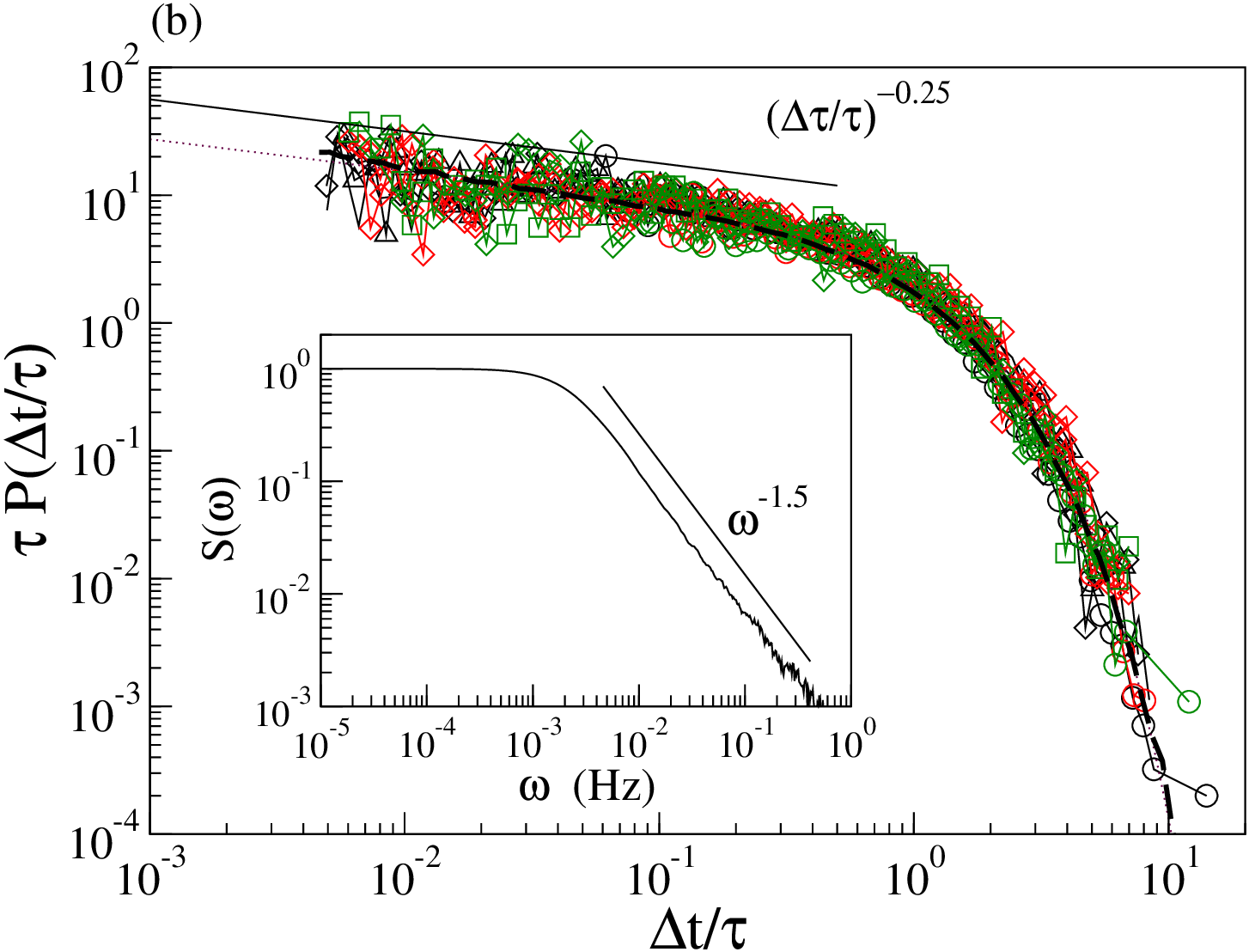}
\end{center}
\caption{\label{fig:intertime} (color online) (a) The distribution of intertimes between two consecutive e-mails of the \junk folders is well described by a Gamma distribution. This is true both for the whole distribution of intertimes, as well as for the distribution of the intertimes sent by given geographical locations (the analysis is restricted to datasets with more than 500 points). Inset: the distribution of the intertimes between regular e-mails. (b) Collapse of the intertime distribution of the \junk folders. The dotted line is a fit to Eq.~\ref{eq:gen_gamma}, the thick dashed line is the result of our model. Inset: the power spectrum of $P(\Delta t)$ for our largest spam dataset.}
\end{figure}

In order to check whether the spamming process could be considered a stationary Poisson process, we have investigated
the probability distribution $P(\Delta t)$ of the intertimes between two consecutive events. For regular e-mails, where correlations are expected and build up when replying to previous e-mails~\cite{Kossinets,Eckmann}, the probability distribution is characterized by a crossover between two power-laws, as shown in Fig.~\ref{fig:intertime}a (inset). For the dataset we have investigated the corresponding exponents are $-0.2$ at short intertimes, and $-1.7$ at large intertimes, but it has been suggested that these exponents may depend on the particular mailbox~\cite{Masuda08}.

Fig~\ref{fig:intertime}a shows the spam intertime probability distributions. We have computed these probability distributions for each \junk folder, as well as for the catalogs obtained from each \junk folder by considering only spams originating from specific geographical locations, China (CHN), European Union (EU) and United States (USA). These different distributions are also shown in  Fig~\ref{fig:intertime}a.
Contrary to the case of regular e-mails, the intertime probability distributions between spams show universal behaviour if time is rescaled by the average rate in each dataset. In fact, $P(t)$ can be expressed as 
\begin{equation} 
\label{eq:scaling}
P(\Delta t) = 1/\tau f(\Delta t/\tau),
\end{equation}
with $f$ an universal function and $\tau$ the mean intertime, as shown by the data collapse obtained by plotting $\tau P(\Delta t)$ versus $\Delta t/\tau$ in Fig~\ref{fig:intertime}b.
The scaling function is well described by a 
generalized gamma distribution,
\begin{equation} 
\label{eq:gen_gamma}
f(\Delta t/\tau) = \alpha \left(\frac{\Delta t}{\tau}\right)^\beta \exp\left[-\left(\frac{\Delta t}{\tau}\right)\right],
\end{equation}
where $\alpha$ is a normalization constant, and $\beta = -0.25$ is found via a regression procedure. The exponent $\beta$ can also be determined from the power spectrum of the intertime distributions, since if $P(\Delta t) \propto \Delta t^\beta$ at small $\Delta t$, then $S(\omega) \propto |\int P(\Delta t) e^{-i\omega \Delta t} d\Delta t|^2 \propto \omega^{-2(\beta+1)}$ at large $\omega$. As $\beta = -0.25$ we expect  $S(\omega) \propto \omega^{-1.5}$, and Fig.~\ref{fig:intertime}(b) (inset) shows that this is actually the case. As $P(\Delta t)$ is not a simple exponential, the spamming process cannot be considered a homogeneous Poisson process.
\begin{figure}[t!]
\begin{center}
\includegraphics*[scale=0.28]{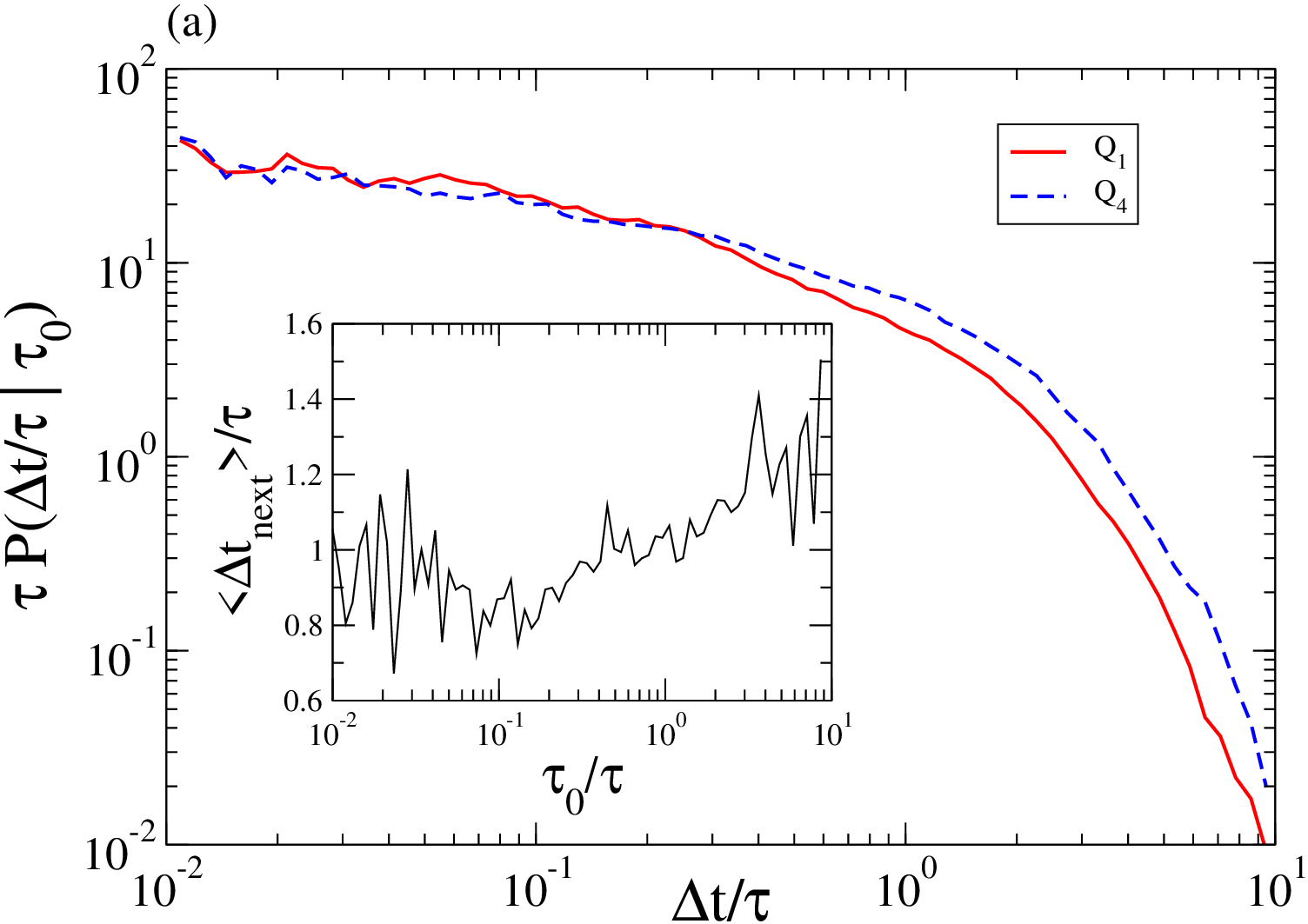}
\includegraphics*[scale=0.28]{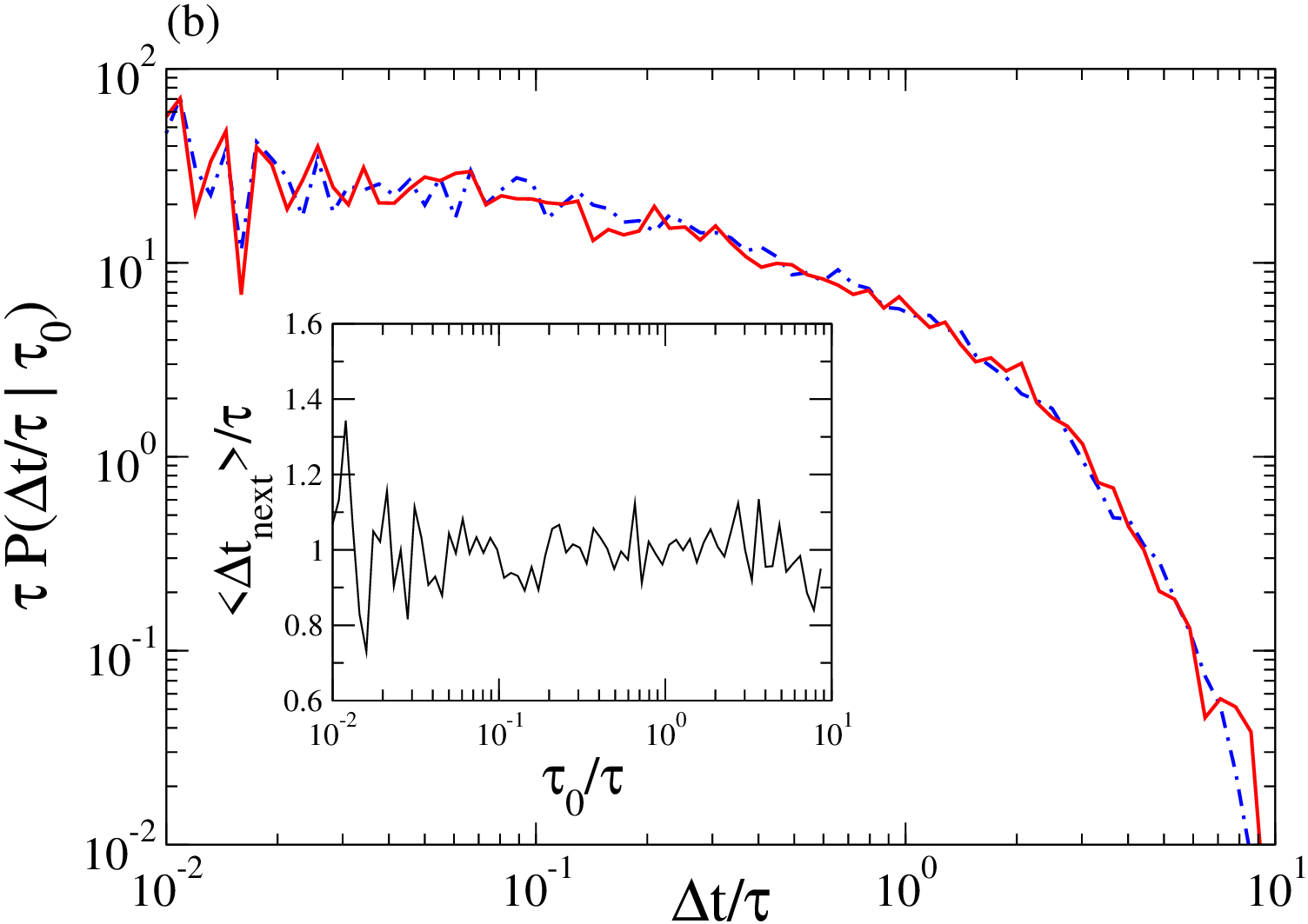}
\includegraphics*[scale=0.28]{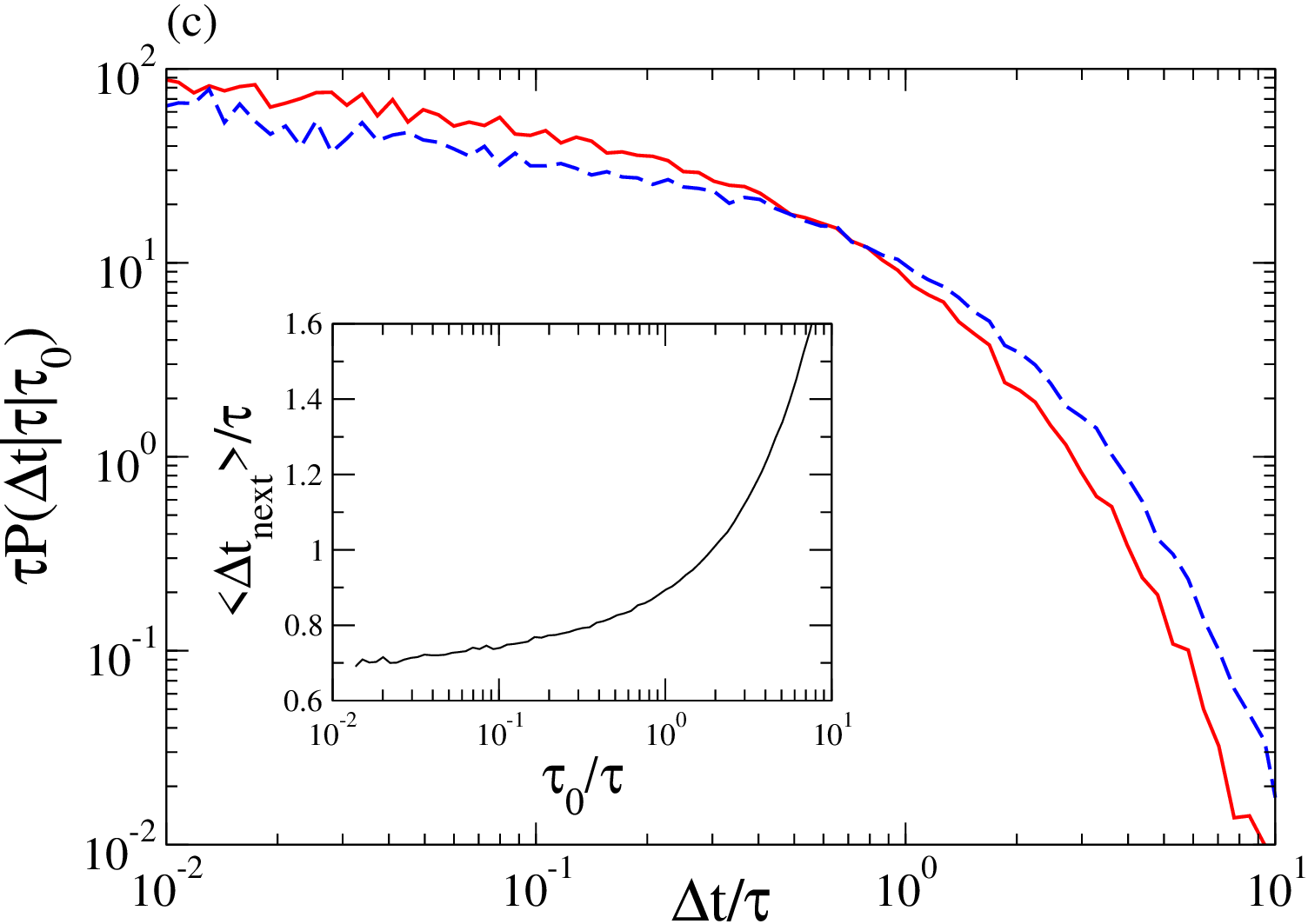}
\end{center}
\caption{\label{fig:conditional} (color online) (a) Conditional probability distribution $P(\Delta t/\tau|\tau_0)$, for $\tau_0 \in Q_1$, 
$\tau_0 \in Q_4$ of the intertimes between spams of the dataset $J_1$. Inset: mean value of the intertime $\Delta t_{next}/\tau$ following an intertime $\tau_0$. (b) The same quantities are shown for the reshuffled catalog, where correlations are absent. (c) The same quantities are shown for the synthetic catalog, where correlations are found.}
\end{figure}

We show now that the non-Poissonian nature of the spamming process is due to the presence of temporal correlations between spams. We have checked that this is actually the case first by computing the mean value $\langle \Delta t_{next}(\tau_0) \rangle$ of the intertimes following an intertime of size $\tau_0$:
\begin{equation}
\langle\Delta t_{next}(\tau_0)\rangle = \frac{\sum_i \Delta t_{i+1} \delta(\Delta t_i,\tau_0)}{\sum_i  \delta(\Delta t_i,\tau_0)},
\end{equation}
where $\delta(\Delta t_i,\tau_0)=1$ if $0.9\tau_0 < \Delta t_i < 1.1 \tau_0$, $0$ otherwise.

In absence of correlations, one expects $\Delta t_{next}(\tau_0)/\tau \simeq 1$. We have determined this quantity both for the original time series of the intertimes $\Delta t_i$ and for the reshuffled time series. These are obtained from the original time series by repeatedly exchanging the values of two randomly chosen intertimes.
Fig.~\ref{fig:conditional}a (inset) shows that the original catalog is characterized by a positive correlation between $\Delta t_{next}/\tau$ and $\tau_0$, which implies that large intertimes are most probably followed by large intertimes. Conversely, for the reshuffled catalog $\Delta t_{next}(\tau_0)/\tau \simeq 1$ (see Fig.~\ref{fig:conditional}b (inset)), indicating absence of correlations.

A further check of the existence of correlations is provided by the study of the conditional probability $P(\Delta t |\tau_0)$, which is the probability of having an intertime $\Delta t$ following an intertime $\Delta t = \tau_0$. To improve the statistics, following~\cite{livina}, we have determined five intertimes $\delta t^*_0 = \min \Delta t_i< \delta t^*_1 <\delta t^*_2 < \delta t^*_3 < \delta t^*_4= \max \Delta t_i$, and defined four subsets $Q_k$, $k = 1,4$, containing respectively the intertimes enclosed between $\delta t^*_{k-1}$ and $\delta t^*_{k}$. The intertimes $\delta t^*_k$, $k=1,2,3$ are chosen in such a way that the four subsets contain the same number of elements.
We have then computed $P_i(\Delta t/\tau|\tau_0)$, $\tau_0 \in Q_i,~ i = 1,2,3,4$. For instance, $P_1$ is the probability distribution of the intertimes (normalized by the mean intertime) following the smallest intertimes, whereas $P_4$ is the probability distribution of the intertimes following the largest intertimes. In absence of correlations these conditional probabilities should all coincide, being equal to the unconditional probability distribution. We show in Fig.~\ref{fig:conditional}a $P_1$ and $P_4$: it is apparent the presence of a systematic difference between the two curves, particularly at large intertimes, where $P_4(\Delta t |\tau_0)~>~P_1(\Delta t/\tau |\tau_0)$. $P_1$ and $P_4$ are also described by Eq.~\ref{eq:scaling}, but the exponents which characterize their power law behavior at short intertimes are different, being equal to $\beta = -0.35$ and $\beta = -0.20$, respectively.
Conversely, for the reshuffled catalog, $P_1$ and $P_4$ do coincide, as shown in Fig.~\ref{fig:conditional}b. From this analysis, we can positively conclude that there exist temporal correlations between spams.

\section{Theoretical model}
These results are very similar to those found in the study of earthquakes~\cite{corral}, where one investigates the statistical features of catalogs registering the time of occurrence of each earthquake, as well as its magnitude. Similar results have been also found in financial markets~\cite{market} and climate~\cite{climate}. The study of earthquakes catalogs has shown that the distribution of intertimes  between earthquakes is described by the same functional form we found for the intertimes between spams~\cite{corral}. Indeed for worldwide seismicity, where the occurrence rate is constant, Corral~\cite{corral} found a Gamma function with a power law initial regime with an exponent $\beta \simeq -0.3$, close to our value. Also the study of the conditional probability distributions~\cite{livina} gives similar results. A deeper understanding of seismic catalogs is however made possible by the fact that earthquakes are characterized by a magnitude, a quantity which has not a counterpart in the case of spams. 
Correlating the magnitude of an earthquake with its occurrence time, it makes possible to clarify that seismic catalogs can be considered as the random superposition of sequences correlated events. Namely, if the first event of the sequence (the mainshock) occurs at time $t = 0$, then the probability that subsequent events (aftershocks) occur at time $t$ is $P(t) \propto t^{-p}$, where $p \simeq 1$ (Omori law).  The sequences are identified considering that the mainshock usually has magnitude greater than subsequent quakes, and actually triggers them.

These considerations suggest that the spam timeseries could also be considered as the random superposition of power law correlated bursts. It is difficult to directly validate this possibility, as the absence of a variable which plays the role of the magnitude makes difficult the identification of the starting event in the burst, even though in some cases bursts are clearly observable in the timeseries, as in Fig.~\ref{fig:rate} (inset). To verify this possibility, we have constructed a synthetic catalog composed by a superposition of $N_s = 10^5$ sequences of $n_b$ events The first element of a sequence occurs at time $t_0$, followed by the others occurring at time $t > t_0$ with probability $P(t-t_0) \propto (t-t_0)^{-p}$, truncated at $t = t_0+10 t^*$, with $t^*$ conventionally set equal to $1$. For each sequence, the starting time $t_0$ is randomly chosen in the interval $[0:t^* N_s]$, and therefore the mean intertime between the beginning of two consecutive sequences is $t^*$.  We have then determined $p$ and $n_b$ fitting the intertime distribution from the model to a Gamma distribution with power low exponent $-0.25$. The best fit, which is shown in Fig.~\ref{fig:intertime}(b), is obtained with $p = 0.8$ and $n_b = 10$. Interestingly, the value of the exponent $p$ is close to the measured value of the Omori exponent for earthquakes. The model exhibits a constant rate, and does also reproduce the intertimes distribution, as shown in Fig.~\ref{fig:intertime}(b), as well as the spam correlations, as shown in Fig.~\ref{fig:conditional}c. These results suggest that, as earthquakes, climate and financial markets, also spams occur in bursts of evenets correlated according to the Omori law.

\section{Conclusions}
It remains open the understanding of the origin of the spam bursts. A step in this direction is given by the analysis of the IP addresses of the computers from which spams are sent. This shows that 1) it never happens that a recipient receives two spams sent by the same IP in a time-window of few hours, and that 2) it occurs that the same spam is sent by the same IP to two different recipients. These evidences, together with the observation that a spammer does not send all of his messages from a single IP, both because it would be too easily discovered, and because it would require too much time due to bandwidth limitations, strongly supports the idea that spammers operate via the use of networks of infected computers, known as botnets~\cite{botnet}. The computational task of sending the e-mails, as well as the bandwidth requirements, is therefore divided among the infected computers. At the present time, there is little information about the structure of botnets, as previous studies have investigated the strategies used to infect a computers, as well as the protocols used by infected computers to communicate.
Nevertheless, our result suggests that botnets are highly dynamical networks, new infected computers being continuously added to replace those that disappear, either because they are switched off, or because their infection is removed via the use of antivirus tools. 
Botnets may also explain the origin of correlations between spams. Indeed, it seems possible that each burst represents the activity of a single spammer. The spammer sends simultaneously more messages to the same user, who receives them at different arrival times, the arrival time of each spam depending on the path followed on the net. As a consequence, the intertimes distribution and the intertimes correlations may result from the topology of the botnet.


\begin{thebibliography}{20}
\bibitem{antispam}
This estimate is provided by a number of companies providing spam filters, such as Messagelabs and Postini.
\end{thebibliography}
\end{document}